\documentclass[letter]{aa}
\usepackage{geometry}
\usepackage{amsmath}
\usepackage{amssymb}
\usepackage{graphicx}
\usepackage{txfonts}
\usepackage{natbib}
\bibpunct{(}{)}{;}{a}{}{,}
\newcommand\cm{\,\rm cm}
\newcommand\s{\,\rm s}

\newcommand\Gyr{\,\rm Gyr}
\newcommand\km{\,\rm km}

\newcommand\kpc{\,\rm kpc}

\newcommand\Msun{\,\rm M_\odot}

\begin{document}

\title{Formation of gaseous arms in barred galaxies with dynamically important magnetic field : 3D MHD simulations}
\titlerunning{Dynamically important magnetic arms in barred galaxies.}
\author{B. Kulesza-\.Zydzik\inst{1}\and
K. Kulpa-Dybe{\l}\inst{1}\and
K. Otmianowska-Mazur\inst{1}\and
G. Kowal\inst{1},\inst{2}\and
M. Soida\inst{1}
}
\authorrunning{B. Kulesza-\.Zydzik et al.}
\offprints{B. Kulesza-\.Zydzik}
\mail{kulesza@oa.uj.edu.pl}
\institute{Astronomical Observatory, Jagiellonian
University, ul Orla 171, 30-244 Krak\'ow, Poland
\and
Department of Astronomy, University of Wisconsin,
475 North Charter Street, Madison, WI 53706, USA}
\date{}

  \abstract
{}
{We present results of three-dimensional nonlinear MHD simulations of a
large-scale magnetic field and its evolution inside a barred galaxy with the back reaction
of the magnetic field on the gas. The model does not consider the dynamo
process. To compare our modeling results with observations, we construct
maps of the high-frequency (Faraday-rotation-free) polarized radio emission on
the basis of simulated magnetic fields. The model accounts for the effects of
projection and the limited resolution of real observations.}
{We performed 3D MHD numerical simulations of barred galaxies and
polarization maps.}
{The main result is that the modeled magnetic field configurations resemble maps
of the polarized intensity observed in barred galaxies. They exhibit polarization
vectors along the bar and arms forming coherent structures similar to the
observed ones. In the paper, we also explain the previously unsolved issue of
discrepancy between the velocity and magnetic field configurations in this type
of galaxies. The dynamical influence of the bar causes gas to form spiral waves
that travel outwards. Each gaseous spiral arm is accompanied by a magnetic
counterpart, which separates and survives in the inter-arm region. Because of a strong compression, shear of
non-axisymmetric bar flows and differential rotation, the total energy of
modeled magnetic field grows constantly, while the azimuthal flux grows slightly
until $0.05\Gyr$ and then saturates.}
  {}


\maketitle


\section{Introduction}\label{intro}
A non-axisymmetric gravitational potential of the bar in barred galaxies
influences strongly the dynamics of stars and the interstellar medium (ISM),
including its magnetic field. The radio polarization observations of
barred galaxies \citep[e.g.,][]{beck-05} show that the magnetic field topology is
far more complicated than the gaseous component would indicate. No physical
process responsible for such a complexity has been found so far. The radio maps
of twenty barred galaxies were presented in \citep{beck-99, beck-02, beck-05} and
\citep{harnett-04}. The main observed features of two of the most spectacular galaxies
NGC~1097 and NGC~1365 \citep{beck-05} can be summarized as follows:
\begin{itemize}
\item the polarized emission (PI) forms ridges coinciding with dust lanes along
leading edges of the bar in the central region of galaxy;
\item the pitch angles of B-vectors vary rapidly in the bar region when
located upstream from dust lanes, producing a depolarization valley
in the telescope beam;
\item the ridges of vanishing polarized intensity are observed in the shear
shock areas;
\item in the outer disk, the polarization vectors form a spiral pattern with the
maximum of emission along the spiral arms but in the inter-arm regions.
\end{itemize}

The issue of magnetic field evolution in barred galaxies was analyzed in several
papers \citep[see][]{otmian-02,otmian-97, moss-98, moss-99, moss-01, moss-07}.
In Moss et al. (2007), three-dimensional (3D)
calculations were applied to study the process of magnetic field dynamo and
gas evolution under the influence of a bar potential. Moreover, they prepared
the global synthetic polarization maps to compare them directly with
the NGC~1365 radio observations. Even though all of the input parameters
have been studied varying each inturn failed to reproduce a configuration of polarization vectors
similar to that observed. The idea of magnetic arms between the gaseous counterparts
was also studied by \cite{rohde-98} using 3D MHD dynamo simulations. The
authors applied a non-uniform distribution of magnetic diffusivity with its higher
values in spiral arms and lower in the inter-arm region. This arrangement was
related to the presence of higher turbulent motions in the spiral arms due to
the presence of star-forming activity there. The assumption of higher magnetic
diffusion in spiral arms, no matter how reasonable, has never been confirmed by
observations. The authors found that it was possible to obtain magnetic arms in
the inter-arm region in the case of resonance of the density pattern speed and a
magnetic drift velocity, thus not for all galaxies. Our nonlinear model of
magnetic field evolution in barred galaxies, which takes into account the back
reaction of magnetic field on the gas, offers an opportunity to understand the
relationship between the large-scale magnetic and gas velocity fields.

Our previous 3D simulations \citep{otmian-02} concerning the large-scale
magnetic field evolution in barred galaxies applied the evolution in velocity
fields predicted by the N-body sticky-particle simulations at that time. It was
impossible to introduce either the back-reaction of magnetic field on the gas
component or the physical mechanism of gaseous spiral waves on the magnetic
field structure. In the present paper, we study the nonlinear evolution of the gas
and magnetic field in a barred galaxy. Our 3D calculations are carried out with
the use of a MHD code written by Kowal (2008).  We do not include the dynamo
process in our MHD equations. The gravitational potential originating in stars
in the form of the bar, halo, and bulge is modeled analytically. The magnetic
field of galaxies is usually
studied using polarized radio-continuum observations, yielding the magnetic
field configuration integrated along the line of sight projected onto the sky
plane and convolved to a certain radio telescope beam. Since our models contain
full three-dimensional magnetic field structures, we analyze them by simulating
the polarization maps and comparing these maps directly with observations.

\section{Methods}\label{method}
\subsection{The initial conditions and input parameters}\label{input}

We investigate the evolution of a barred galaxy solving the resistive MHD
equations \citep{landau}. We apply an isothermal equation of state $\mbox{\boldmath $p_{gas} = \rho c_{s}^2$}$, 
where $c_s=5$ km s$^{-1}$ is a constant isothermal speed of sound. In our model,
we assume that the initial gas density distribution is of the form:
\begin{equation}
\label{eq:dens_d}
\rho = \rho_0\exp\left(-\frac{|z|}{h}\right)Q(r),
\end{equation}
where $\rho_0=1$ H $\cm^{-3}$ is the initial gas density of the neutral hydrogen
in the galactic mid-plane, $h=0.3\kpc$ is the gas scale-height, and function $Q(r)$
is the truncation factor that reduces the density to small values at $r>R_{max}$,
$R_{max}=9$ kpc being the galaxy radius. We assume the azimuthal initial magnetic
field configuration ($B_z=0$, $B_r=0$, $B_{\varphi}(z,r)$). Its distribution
depends strictly on the gas density distribution via the condition $\beta=\frac{p_{gas}}{p_{mag}}$, 
where $\beta=10.0$ at the beginning (determining the initial value of magnetic
field to be $B_0=1.0 \mu G$ in the galactic mid-plane). We use the
turbulent diffusion coefficient for the interstellar gas $\eta=3\cdot10^{25}\cm^2\s^{-1}$ \citep{Lesch93}.

The gravitational potential of stars is introduced analytically. Our galaxy is
initially composed of three components: the large and massive halo, the central
bulge and the rotating disk of stars. The circular speed of the gas is:
\begin{eqnarray}
\label{eq:rotation}
V_{rot}^2 & = & \frac{G M_d r^2}{\left(a_d+\sqrt{a_{d}^2+r^2}\right)^2 \sqrt{a_{d}^2+r^2}}+\nonumber\\
&+&GM_b(r^2+a_{b}^2)^{-3\slash 2}+GM_h(r^2+a_{h}^2)^{-3 \slash 2}+\nonumber\\
 &+& \frac{c_{s}^{2}(1+\beta) r}{\beta Q} \frac{\partial Q}{\partial r} ,
\end{eqnarray}
where $M_d=10^{10}\Msun$ is the mass of the disk of stars, $a_d=0.2$ kpc is its
characteristic scale-length, and both $M_b=10^{10}\Msun$, and $a_{b}=2$~kpc refer to the
bulge and $M_h=10^{11}\Msun$, and $a_{h}=20$~kpc refer to the halo. The bar
component grows gradually in time (from $t_{beg}=0.1 Gyr$ to $t_{end}=0.4 Gyr$),
changing the gravitational potential of the galaxy. The bar potential is
described by the second-order Ferrers ellipsoid \citep{ferrers-1877}  with
semi-axes $a=4kpc$, $b=2\kpc$, and $c=2\kpc$. To conserve the total mass of
the galaxy, we reduce the bulge mass, such that $M_{bar}(t)+M_b(t)=M$, where $M$ is constant
during the calculations. The bar angular velocity $\Omega_{bar}$ is set to be
$30\km\s^{-1}\kpc^{-1}$. This determines the value of the corotation radius to
be $R_{cor}=4.46\kpc$.

We use a higher-order shock-capturing Godunov-type scheme based on the
essentially non-oscillatory (ENO) spacial reconstruction, a Harten-Lax-van-Leer
(HLL) approximate Riemann-solver and Runge-Kutta (RK) time integration
\citep[e.g.]{delzanna03} to solve isothermal non-ideal MHD equations.
We evolve the induction equation using the constrained transport (CT) method 
\citep{Evans} to ensure that magnetic field divergence vanishes everywhere at all times.

The computational domain extends from $-10\kpc$ to $10\kpc$ in the $x$~and $y$
directions, and from $-2.5\kpc$ to $2.5\kpc$ in the $z$ direction with the grid 
size $n_x=n_y=512$, $n_z=129$.

\subsection{Stokes parameters}

To compare our results with observed properties of barred galaxies, we
constructed synthetic radio-polarization maps (face on) at selected time steps. We
calculated the maps by integrating Stokes parameters of the synchrotron emissivities I,
Q, and U along the line of sight and finally convolving them with an assumed
Gaussian beam. As a result, we generated polarization maps of the intensity and
the angle of polarization $B$-vectors. We assumed that the distribution of
relativistic electrons change radially and vertically as Gaussian functions. The
truncation radius and scale height were respectively equal to $9$~kpc and $1$~kpc.
The adopted intrinsic polarization degree of synchrotron emission is $70\%$ and
the energy spectral index of the relativistic electrons $\gamma$ is $2.8$.

\section{Results}\label{results}

\subsection{Velocity structures and magnetic field amplification}

\begin{figure}[t]\begin{center}
{\includegraphics[width=0.8\columnwidth]{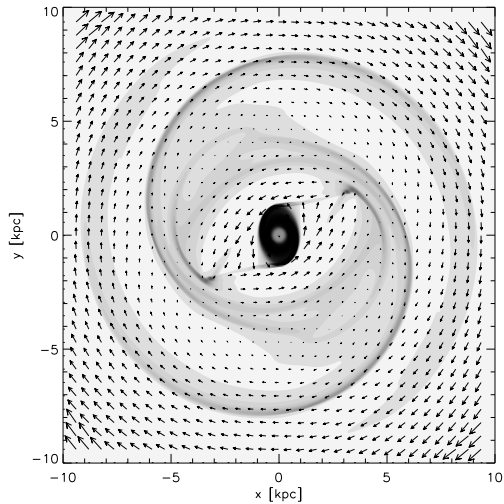}}
\caption{Velocity field in the rest frame of the bar (vectors) and the gas density distribution (grey-scale, dark means high values of density) along the $z=0$ plane at time $t=0.64\Gyr$.}
\label{fig:vel_field}
\end{center}\end{figure}

\begin{figure}[!t]\begin{center}
\includegraphics[width=0.8\columnwidth]{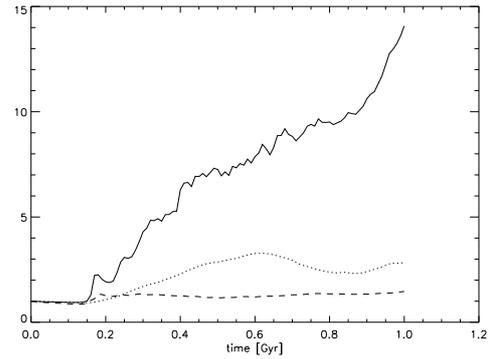}
\caption{Time dependence of the total magnetic field energy $ B^2 $ (solid line)
and the mean $ B_\phi $ flux in the galactic mid-plane (dashed). The third plot (dotted) presents the total 
magnetic field energy without taking into account the innermost bar region. All values are
normalized to their initial values.}
\label{fig:mf_fluxes}
\end{center}\end{figure}

Figure~\ref{fig:vel_field} shows the gas density distribution (grey-scale plot)
and velocity field (in a reference frame corotating with the bar) in the galactic
mid-plane at time $t=0.64\Gyr$. The bar's major axis is inclined by $20\degr$
to the x-axis. The high density regions appearing from the innermost region of
the galaxy indicate the leading edge of the bar. Density enhancements are
visible at the ends of the bar and in the spiral arms generated in the galactic
disk. The bar strongly disturbs the gas velocity field. Rapid variations in the
velocity vector direction can be seen in the innermost region of the bar. The
gas flows out of the center just before it passes the shock, and then turns
back toward the center along the leading edge of the bar. Gas motions generate
shocks that strongly influence the magnetic field structure. The bar is
gradually introduced to the model from $t_{beg}=0.1\Gyr$ to $t_{end}=0.4\Gyr$.
During this time, as we can see in (Fig.~\ref{fig:mf_fluxes}), the mean value of the
$B_{\phi}$ flux in the galactic mid-plane increases slightly and then saturates.
No significant enhancement in the $B_{\phi}$ flux can be seen. The total magnetic
field energy ($B^2$) increases as the galaxy evolves, because of an increase in
the magnetic field complexity, which is in turn the result of a complex 
velocity field global structure. In our model, the most significant part of the growth in
the magnetic field energy originates in the combined effects of shearing motions
and compression present in the innermost bar region.

\subsection{The polarization maps}

\begin{figure*}\begin{center} 
 {\includegraphics[width=0.6\columnwidth]{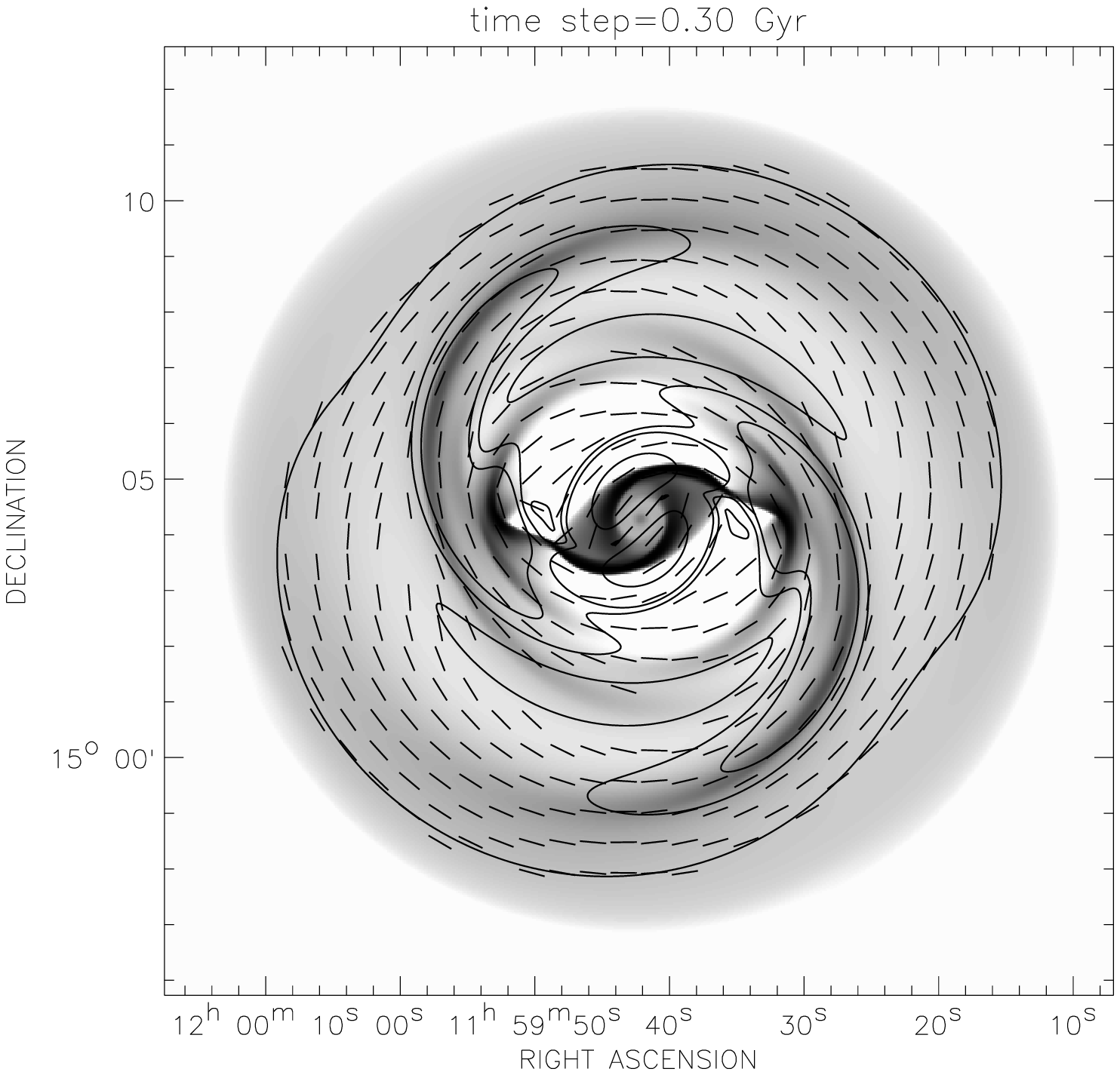}\includegraphics[width=0.6\columnwidth]{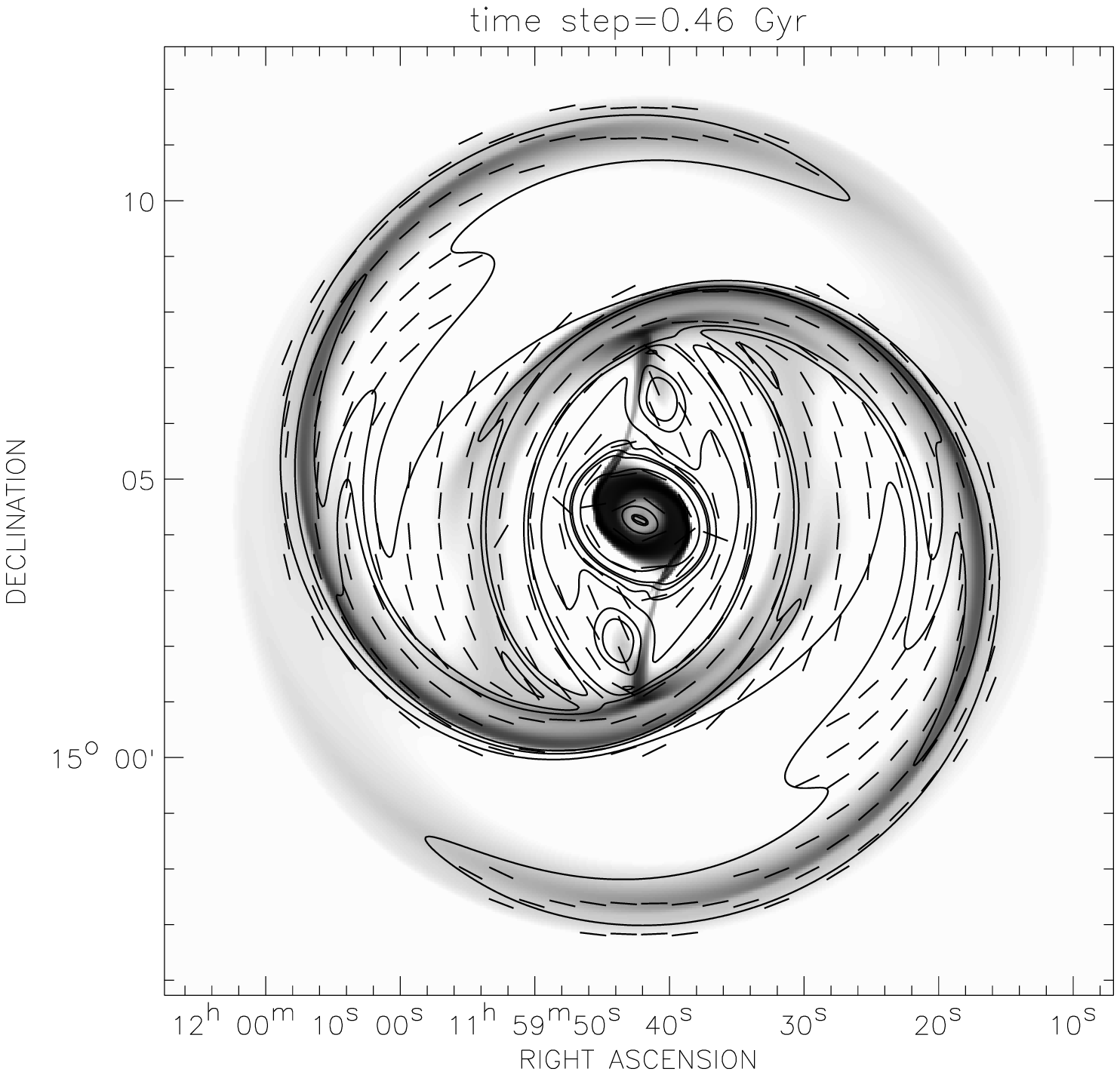}\includegraphics[width=0.6\columnwidth]{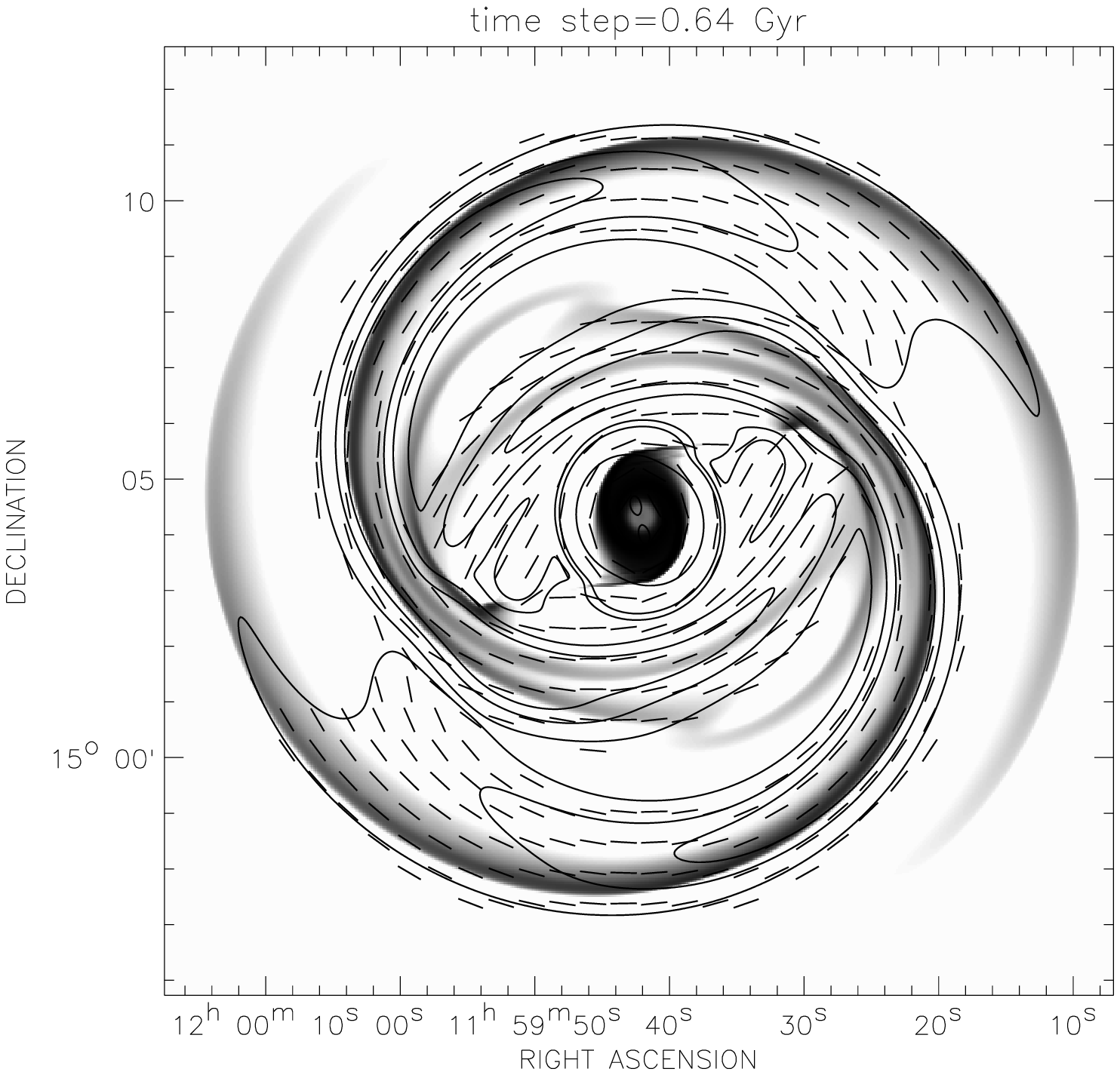}}
 \caption{Face-on polarization maps at $\lambda=6.2\cm$ at selected times steps ($0.3\Gyr$ left, $0.46\Gyr$ middle, $0.64\Gyr$ 
right): polarized intensity (contours), polarization angle (vectors) and column density (grey-scale). All maps have been smoothed 
down to the resolution $40''$. The black color represents the regions with the highest density.} \label{fig:polar}
\end{center}\end{figure*}

Figure~\ref{fig:polar} shows the distributions of
polarization angle (vectors) and polarized intensity (contours) superimposed
on the column density (grey-scale) at time t=0.64Gyr. To illustrate the evolution of the magnetic field
configuration in our model, we present results for three crucial time steps:
0.3~Gyr (Fig.~3 left), 0.46 ~Gyr (Fig.~3 in the middle), and 0.64~Gyr (Fig.~3
right). As mentioned above, in the beginning the modeled barred galaxy is in
magneto-hydrostatic equilibrium. After almost one and a half rotation periods
($t_1=0.30\Gyr$), the density distribution diverges significantly from its
initial form. In the inner part of the disk, where the bar is present, we find
the highest gaseous density and magnetic field intensity regions.

The compression of both the gas and magnetic field is also visible in the outer part
of the disk in the form of spiral arms. All of these gaseous structures are created
by the dynamical influence of the non-axisymmetric gravitational potential
of the bar. This potential also changes the initial azimuthal
configuration of the magnetic field. First, the magnetic field follows the
gas distribution, as can easily be seen for time step $t_1=0.30\Gyr$
in Fig.~\ref{fig:polar} (left), where the magnetic field strength maxima are aligned
with the gaseous ones. However, at later time-steps, the magnetic arms begin to
detach themselves from the gaseous spirals and drift into the inter-arm regions preserving
similar to the gaseous arms pitch angles (see $t_2=0.46\Gyr$
Fig.~\ref{fig:polar}, middle). At $t_3=0.64\Gyr$ (Fig.~\ref{fig:polar}, right),
we can observe that our magnetic arms are distributed between density wave
structures. The magnetic arms do not corotate with gaseous spiral structure, but
they follow the general gas motion in the disk, which has a slightly lower angular
velocity. The last map (Fig.~\ref{fig:polar}, right) also shows the
depolarization regions present in the bar. They are the effect of so-called beam
depolarization, where magnetic field lines both along the bar and in the
inter-arm regions converge with large angles within small, one-beam-size
areas. Similar results were also obtained by \cite{otmian-02}.

\subsection{Magnetic pitch angles}

\begin{figure}[h]\begin{center} 
 {\includegraphics[width=1\columnwidth]{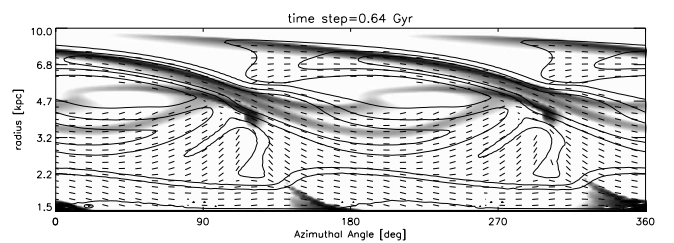}}
 \caption{The distribution of polarized intensity and $B$-vector orientation superimposed
 on the gas density at $\lambda=6.2\cm$ in the frame of azimuthal angle in the disk and
 the logarithm of distance from the center. The map has been convolved to the resolution $40''$.}
  \label{fig:phase}
\end{center}\end{figure}

To analyze and compare the pitch angles of the magnetic and gaseous arms, we project
our results onto the polar coordinate space of azimuthal angle in the disk and $\ln(r)$,
where $r$ is the galactocentric distance. In this case, the logarithmic spiral is
represented by a straight line inclined by its pitch angle. The grey plot of the gas
density (integrated along the line of sight) and superimposed contours of polarized
intensity and $B$-vectors at a time $0.64\Gyr$ are shown in Fig.~\ref{fig:phase}.
The values of the pitch angle of the magnetic arms are similar to those of the gaseous
ones, i.e., they are of order of $5\degr$. We also estimated the mean pitch angle
(averaged over the azimuthal angle and radius in the plane of galaxy) to be $3\degr$.
Since we do not apply dynamo effects, the aforementioned pitch angle values are rather
low, slightly smaller than usually observed in spiral galaxies \citep[see][]{beck-93}.
However, between the radii $r=4.1\kpc$ and $r=4.5\kpc$, the mean pitch angles are much
higher and reach $24\degr$.

In the inner part of the disk (between the bar and spiral arms), the $B$-vectors of the
polarized intensity change their pitch angles rapidly (Fig~\ref{fig:phase}, see $\phi$
around 100 and 300, and ln r between 2.2 and 3.2). This is caused by the strong flow
of the gas between the end of the bar and the magnetic spiral arms, which produces valleys
of depolarization within a telescope beam. This bears some resemblance to the observations
of NGC~1356 \citep{beck-05} and NGC~3627 \citep{soi01} , even though we applied different
rotation curves.

\section{Discussion \& Conclusions}\label{discussion}

For the first time, we have simulated the evolution of large-scale galactic magnetic fields
in barred galaxies using 3D MHD nonlinear numerical simulations without a dynamo process.
The physical processes modeled and observed in our simulations can be summarized as follows.
First, the dynamical influence of the bar causes the gas to form spiral waves. Later on,
both the shear and compression in the arms cause the magnetic arms to be aligned with the
gaseous structures. Since the pattern of the gaseous spiral arms travels faster than the disk
itself, especially beyond the corotation radius, the gaseous structures move further out,
while the magnetic arms remain in the interarm regions. This results in the magnetic and spiral
arms having similar pitch angles.

We found that:
\begin{itemize}
\item[1.]{Magnetic arms develop inside the gaseous ones, but gradually diverge away from the
desnity waves. The magnetic arms are consequently shifted toward the interarm regions, in
qualitative agreement with observations.}
\item[2.]{The above effect is also responsible for prducing the depolarized valleys in the bar
region, which also agrees with observations. }
\item[3.]{The magnetic pitch angles of the synthetic magnetic arms are similar to the pitch
angles measured in spiral gaseous structures, which also agrees with observations.}
\end{itemize}

This process could also be responsibile for the shifting of the magnetic arms to the interarm regions in spiral galaxies.

\begin{acknowledgements}
The authors express their gratitude to R. Beck  and M. Urbanik for helpful comments.
This work was supported by Polish Ministry of Science and Higher Education through grants:
92/N-ASTROSIM/2008/0 and 3033/B/H03/2008/35. Presented computations have been performed on the GALERA supercomputer in TASK Academic 
Computer Centre in Gdansk.
\end{acknowledgements}


\end{document}